\documentclass[twocolumn,showpacs,preprintnumbers,amsmath,amssymb]{revtex4}

\usepackage{graphicx}

\begin{document}

\title{QED$_3$ on a space-time lattice: compact versus noncompact formulation}

\author{R. Fiore}
 \email{fiore@cs.infn.it}
\affiliation{Dipartimento di Fisica, Universit\`a della Calabria
             \& Istituto Nazionale di Fisica Nucleare, Gruppo collegato di
             Cosenza, I-87036 Arcavacata di Rende, Cosenza, Italy}

\author{P. Giudice}
 \email{giudice@cs.infn.it}
\affiliation{Dipartimento di Fisica, Universit\`a della Calabria
             \& Istituto Nazionale di Fisica Nucleare, Gruppo collegato di
             Cosenza, I-87036 Arcavacata di Rende, Cosenza, Italy}

\author{D. Giuliano}
 \email{giuliano@cs.infn.it}
\affiliation{Dipartimento di Fisica, Universit\`a della Calabria
             \& Istituto Nazionale di Fisica Nucleare, Gruppo collegato di
             Cosenza, I-87036 Arcavacata di Rende, Cosenza, Italy}

\author{D. Marmottini}
 \email{marmotti@cs.infn.it}
\affiliation{Dipartimento di Fisica, Universit\`a della Calabria
             \& Istituto Nazionale di Fisica Nucleare, Gruppo collegato di
             Cosenza, I-87036 Arcavacata di Rende, Cosenza, Italy}

\author{A. Papa}
 \email{papa@cs.infn.it}
\affiliation{Dipartimento di Fisica, Universit\`a della Calabria
             \& Istituto Nazionale di Fisica Nucleare, Gruppo collegato di
             Cosenza, I-87036 Arcavacata di Rende, Cosenza, Italy}

\author{P. Sodano}
 \email{pasquale.sodano@pg.infn.it}
\affiliation{Dipartimento di Fisica,~Universit\`a di Perugia
             \& Istituto Nazionale di Fisica Nucleare,\\
	     via A.~Pascoli, I-06100  Perugia, Italy}

\date{\today}

\begin{abstract}
We study quantum electrodynamics in a (2+1)-dimensional space-time
with two flavors of dynamical fermions by numerical simulations on
the lattice. We discretize the theory using both the compact and the
noncompact formulations and analyze the behavior of the chiral
condensate and of the monopole density in the finite lattice regime
as well as in the continuum limit. By comparing the results obtained
with the two approaches, we draw some conclusions about the possible
equivalence of the two lattice formulations in the continuum limit.
\end{abstract}


\maketitle

\section{\label{sec:1} Introduction}

Quantum electrodynamics in 2+1 dimensions (QED$_3$) is interesting
as a toy model for investigating the mechanism of confinement in
gauge theories~\cite{Po}, and as an effective description of
low-dimensional, correlated, electronic condensed matter systems,
like spin systems~\cite{us,senthil}, or high-$T_c$ superconductors~\cite{MaPa}. 
While the compact formulation of QED$_3$ appears to be
more suitable for studying the mechanism of confinement, both 
compact~\cite{AM} and noncompact formulations arise in condensed matter
systems. Our paper aims to elucidate some aspects of the
relationship between these two formulations of QED$_3$ on the
lattice.

Polyakov showed that compact QED$_3$ without fermion degrees of
freedom is always confining~\cite{Po}. Any pair of test
electric charge and anti-charge is confined by a linear potential,
as an effect of proliferation of instantons, which are magnetic
monopole solutions in three dimensions. The plasma of such monopoles
is what is responsible for confinement of electrically charged
particles. If compact QED$_3$ is coupled to matter fields it has
been argued~\cite{deconf} that the interaction between monopoles
could turn from $1/x$ to $-\ln{(x)}$ at large distances $x$, so that
the deconfined phase may become stable at low temperature. The issue
of the existence of a confinement-deconfinement transition in
QED$_3$ at $T=0$ is still controversial, as it has also been
proposed that compact QED$_3$ with massless fermions is always in
the confined phase~\cite{HeSe,AzLu}; also, in the limit of large flavor number,
it has been argued that monopoles should not play any role in the
confinement mechanism~\cite{Herm}. At finite temperature, parity
invariant QED$_3$ coupled with fermionic matter undergoes a 
Berezinsky-Kosterlitz-Thouless transition to a deconfined phase~\cite{griseso}.

The issue of charge confinement in $2+1$ dimensional gauge models
comes out to be relevant in the context of quantum phase
transitions, as well. Indeed, recently it has been proposed that
phenomena similar to deconfinement in high energy physics might
appear in planar correlated systems, driven to a quantum (that is,
zero-temperature) phase transition between an antiferromagnetically
ordered (Ne\'el) phase, and a phase with no order by continuous
symmetry breaking~\cite{us,senthil}. The most suitable candidate for
a theoretical description of the system near the quantum critical
point is a planar gauge theory, either with Fermionic matter~\cite{us}, 
or with Bosonic matter~\cite{senthil}.

At finite $T$ noncompact QED$_3$ comes about to be relevant in the
analysis of the pseudo gap phase~\cite{pseudo} of cuprates. This
phase arises from the fact that, upon doping the cuprate, a gap
opens at some temperature $T^\star$ which is quite larger than the
critical temperature $T_C$ for the onset of superconductivity. Both
temperatures $T^\star$ and $T_C$ are doping dependent quantities and
the gap is strongly dependent upon the direction in momentum space,
since it exhibits $d$-wave symmetry~\cite{TsuKi}. 

\begin{figure}[ht]
\includegraphics[width=2.8in]{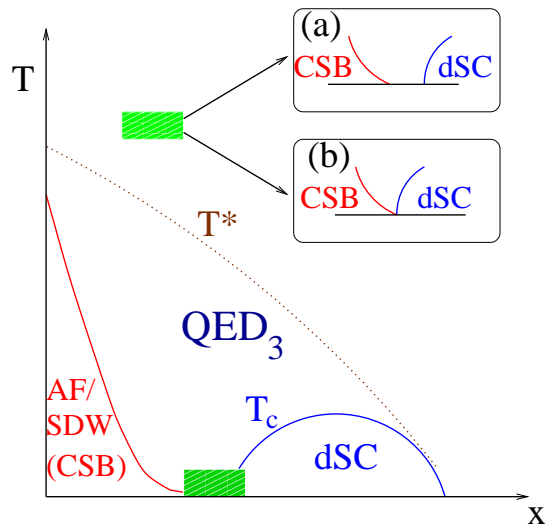}
\caption{Phase diagram in the $(x,T)$ plane~\cite{Fra};
here $x$ represents the doping and $T$ the temperature.}
\label{phase_diagram}
\end{figure}

In Fig.~\ref{phase_diagram} we report the phase diagram of high-$T_c$
cuprates.
For small-$x$ phase is characterized~\cite{HaTh} by an insulating
antiferromagnet (AF); by increasing $x$, this phase evolves into a spin
density wave (SDW), that is a weak antiferromagnet. The pseudo gap phase is
located between this phase and the $d$-wave superconducting (dSC) one.

The effective theory of the pseudo gap phase~\cite{pseudo} turns out to be
QED$_3$~\cite{MaPa,Fra,Her}, with spatial anisotropies in the
covariant derivatives, that is with different values for the Fermi
and the Gap velocities~\cite{HaTh}, and with Fermionic matter given
by spin-$1/2$ chargeless excitations of the superconducting state
(spinons). These excitations are minimally coupled to a massless
gauge field, which arises from the fluctuating topological defects
in the superconducting phase. The SDW order parameter is
identified with the order parameter for chiral symmetry breaking (CSB) in
the gauge theory, that is, $\langle \bar{\psi} \psi \rangle$~\cite{Her}. 
There can be two possibilities; if $\langle \bar{\psi}
\psi \rangle$ is different from zero, then the $d$-wave
superconducting phase is connected to the spin density wave
one (see Fig.~\ref{phase_diagram} case b); otherwise the two
phases are separated at $T=0$ by the pseudo gap phase (see
Fig.~\ref{phase_diagram} case a).

Confinement and chiral symmetry breaking go essentially
together as strong coupling phenomena in gauge theories; while
confinement is an observed property of the strong interactions and
it is an unproven, but widely believed feature of non-abelian gauge
theories in four space-time dimensions, chiral symmetry is only an
approximate symmetry of particle physics, since the up and down
quarks are light but not massless. Central to our understanding of
CSB is the existence of a critical coupling: when fermions have a
sufficiently strong attractive interaction there is a pairing
instability and the ensuing condensate breaks some of the flavor
symmetries, generate quark masses, and represents chiral symmetry in
the Nambu-Goldstone mode~\cite{namgot,modernnamgot}. The issue of a
critical coupling has been widely investigated in 2+1 dimensional
gauge theories~\cite{cc1,cc2,cc3}. Typically, the dimensionless
expansion parameter is $1/N_{f}$. Using the Schwinger-Dyson
equations~\cite{cc1} or a current algebra approach~\cite{diaseso}
for QED$_3$ and QCD$_3$ one finds that there is a critical number of
flavors, $N_{f,c}$, such that only for $N_{f}$ lesser than $N_{f,c}$
chiral symmetry is broken; for $N_{f}$ bigger than $N_{f,c}$
chirality is unbroken and quarks remain massless. For QED$_3$ this
result has been the subject of some debate~\cite{cc1,cc2,cc4,cc5,ApWi,ApCoSc,
Fischer:2004nq}; there are, however, numerical simulations~\cite{simulflav,HaKoSt,
HaKoScSt} of QED$_3$, which find an $N_{f,c}$ remarkably close to the results
reported in Ref.~\cite{cc1}.

Even if far from the scaling regime, strong coupling gauge theories
on the lattice provide interesting clues on the issue of CSB. In
fact, one can show that, in the strong coupling limit, a Hamiltonian
with $N_{c}$ colors of fermions and $N_{f}/2$ lattice flavors of
staggered fermions is effectively a $U(N_{f}/2)$ quantum
antiferromagnet with representations determined by $N_{c}$ and
$N_{f}$~\cite{strongcoupling}. CSB is then
associated~\cite{strongcoupling} either to the formation of a $U(1)$
commensurate charge density wave or of a $SU(N_f/2)$ spin density
wave, i.e. to the formation of Ne\'el order. Quantum antiferromagnets
with the representations considered in Ref.~\cite{strongcoupling} have
been analyzed in Ref.~\cite{reasac} where it was found that, for small
enough $N_f$, the ground state is ordered. Also, when $N_f$ is
increased there is a phase transition, for $N_f \sim N_c$, to a
disordered state. In this picture, the large $N_c$ limit is the
classical limit where Ne\'el order is favored and the small $N_c$ and
large $N_f$ limit are where fluctuations are large and disordered
ground states are favored.

We shall not try to ascertain in this paper the critical number of
flavours $N_{f,c}$. Here, we shall analyze the relationship between monopole 
density and fermion mass and compare the results obtained for the compact and 
noncompact lattice formulation of this gauge model.
In particular, we revisit the 
analysis of Fiebig and Woloshyn of Refs.~\cite{FiWo1,FiWo2}, where the dynamic 
equivalence
between the two formulations of (isotropic) QED$_3$ is claimed to be
valid in the finite lattice regime. In this paper we shall extend the
comparison to the continuum limit, following the same approach as in
Refs.~\cite{FiWo1,FiWo2}: namely we shall analyze the behavior of
the chiral condensate and of the monopole density as the continuum
limit is reached.

In Section II we describe the model and its properties both in the
continuum and on the lattice. Moreover, the method for detecting
monopoles on the lattice is illustrated.

In Section III a description of both compact and noncompact
formulations of QED$_3$ is given.

In Section IV we present our numerical result for the chiral
condensate and the monopole density in the region in which the
continuum limit is reached. Then, we compare our results with those
of Fiebig and Woloshyn~\cite{FiWo1,FiWo2}.

Section V is devoted to conclusions.

\section{\label{sec:2} The model and its properties}

The continuum Lagrangian density describing QED$_3$ is given in
Minkowski metric~\cite{Ap} by
\begin{equation}
{\cal L} =-\frac{1}{4} F_{\mu \nu}^2+\overline\psi_i iD_\mu \gamma^\mu \psi_i -
 m_0\overline\psi_i \psi_i \;,
\end{equation}
where $D_\mu=\partial_\mu-ieA_\mu$, $F_{\mu \nu}$ is the field
strength and the fermions $\psi_i$  ($i=1, \dots, N_f$) are
4-component spinors. Since QED$_3$ is a super-renormalizable theory,
dim$[e]=+1/2$, the coupling does not display any energy dependence.
One may define three $4 \times 4$ Dirac matrices
\[
\gamma^0=\left(
         \begin{array}{cc}
         \sigma_3 & 0 \\
         0        & -\sigma_3
         \end{array}
         \right)\;, \hspace{1cm}
\gamma^1=\left(
         \begin{array}{cc}
         i\sigma_1 & 0 \\
         0        & -i\sigma_1
         \end{array}
         \right)\;,
\]
\begin{equation}
\gamma^2=\left(
         \begin{array}{cc}
         i\sigma_2 & 0 \\
         0        & -i\sigma_2
         \end{array}
         \right)\;,
\end{equation}
and two more matrices  anticommuting with them: namely
\begin{equation}
\gamma^3=i\left(
         \begin{array}{cc}
         0 & 1 \\
         1 & 0
         \end{array}
         \right)\;,\hspace{1cm}
     \gamma^5=i\left(
         \begin{array}{cc}
         0 & 1 \\
         -1 & 0
         \end{array}
         \right)\;.
\end{equation}
The massless theory will therefore be invariant under the chiral
transformations
\begin{equation}
\psi \rightarrow e^{i \alpha \gamma^3} \psi \;, \qquad
\psi \rightarrow e^{i \beta \gamma^5} \psi \;.
\end{equation}
If one writes a 4-component spinor as $ \psi=\left(
\begin{array}{c}
\psi_1  \\
\psi_2
\end{array}
\right),
$
the mass term becomes
$$
m\overline\psi\psi=m\psi_1^\dagger\sigma_3\psi_1-m\psi_2^\dagger\sigma_3
\psi_2 \;.
$$
Since in three dimensions the parity transformation reads
\begin{eqnarray}
\psi_1(x_0,x_1,x_2) &\rightarrow & \sigma_1\psi_2(x_0,-x_1,x_2) \;,
\nonumber \\
\psi_2(x_0,x_1,x_2) &\rightarrow & \sigma_1\psi_1(x_0,-x_1,x_2) \;,
\end{eqnarray}
then $m\overline\psi \psi$ is parity conserving.


The lattice Euclidean action~\cite{AlFaHaKoMo,HaKoSt} using
staggered fermion fields $\overline\chi,\chi$, is given by
\begin{equation}
\label{action}
 S=S_G+\sum_{i=1}^{N} \sum_{n,m}  \overline\chi_i(n)
M_{n,m} \chi_i(m)\;,
\end{equation}
where $S_G$ is the gauge field action and the fermion matrix is given by
\begin{eqnarray}
&M_{n,m}[U]& \\
&=\sum_{\nu=1,2,3} \frac{\eta_{\nu}(n)}{2} \left\{ [U_{\nu}(n)
    ]  \delta_{m,n+\hat{\nu}}-
    [U_{\nu}^{\dagger}(m)] \delta_{m,n-\hat{\nu}} \right\} \;.& \nonumber
\end{eqnarray}

The action (\ref{action}) allows to simulate $N=1,2$ flavours of
staggered fermions corresponding to $N_f=2,4$ flavours of
4-component fermions $\psi$~\cite{BuBu}. $S_G$ is different for the
compact and noncompact formulation of QED$_3$.

For the compact formulation one has
\begin{equation}
S_G[U]= \beta \sum_{n,\mu < \nu}{ \left[ 1 - \frac{1}{2} \left( U_{\mu \nu}
 (n)+ U_{\mu \nu}^{\dagger}(n) \right) \right]}\;,
\end{equation}
where $U_{\mu \nu}(n)$ is the ``plaquette variable'' and
$\beta=1/(e^2 a)$, $a$ being the lattice spacing. Instead, in the
noncompact formulation one has
\begin{equation}
S_G[\alpha]= \frac{\beta}{2} \sum_{n,\mu < \nu} F_{\mu \nu}(n)F_{\mu \nu}(n)\;,
\end{equation}
where
\begin{equation}
F_{\mu \nu}(n)=  \{ \alpha_{\nu}(n+\hat{\mu})-
  \alpha_{\nu}(n) \} - \{ \alpha_{\mu}(n+\hat{\nu})-
  \alpha_{\mu}(n) \}
\end{equation}
and $\alpha_{\mu}(n)$ is the phase of the ``link variable''
$U_{\mu}(n)=e^{i \alpha_{\mu}(n)}$, related to gauge field by
$\alpha_{\mu}(n)=a e A_{\mu}(n)$.


Monopoles are detected in the lattice using the method given by
DeGrand and Toussaint~\cite{DeTo}: due to the Gauss's law, the total
magnetic flux emanating from a closed surface allows to determine if
the surface encloses a monopole. The monopole density is defined by half of the 
total number of monopoles and antimonopoles divided by the
number of elementary cubes in the lattice. We apply this definition
for both the compact and the noncompact formulations of the theory,
although some caution should be used in this respect. Indeed,
monopoles are classical solutions of the theory with finite action
only for compact QED$_3$, where they are known to play a relevant
role. In the noncompact formulation of QED$_3$ they are not
classical solutions, but they could give a contribution to the
Feynman path integral owing to the periodic structure of the
fermionic sector~\cite{HaWe}.

\section{\label{sec:3} Compact versus noncompact formulation}

In order to investigate the onset of the continuum physics, it is
convenient to consider a dimensionless observable and to evaluate it
from the lattice for increasing $\beta$ until it reaches a plateau.
Such an observable can be taken to be $\beta^2
\langle\overline\chi\chi\rangle$, which is expected to become
constant in the continuum ($\beta \rightarrow \infty$)
limit~\cite{simulflav,DaKoKo2}. Numerical simulations show two
regimes: for $\beta$ larger than a certain value, the theory is in
the continuum limit (flat dependence of a dimensionless observable
from $\beta$), otherwise the system is in a phase with finite
lattice spacing. In the former regime, the theory describes
continuum physics, in the latter one it is appropriate to describe a
lattice condensed-matter-like system.

There are a couple of papers by Fiebig and Woloshyn in which the two
formulations are compared in the finite lattice
regime~\cite{FiWo1,FiWo2}. In these papers the $\beta$-dependence of
the chiral condensate and of the monopole density for lattice
QED$_3$ with $N_f=0$ and $N_f=2$ are analyzed for both compact and
noncompact formulations in the finite lattice regime.

\begin{figure}[ht]
\includegraphics[width=8cm,bb=5 5 596 790,clip]{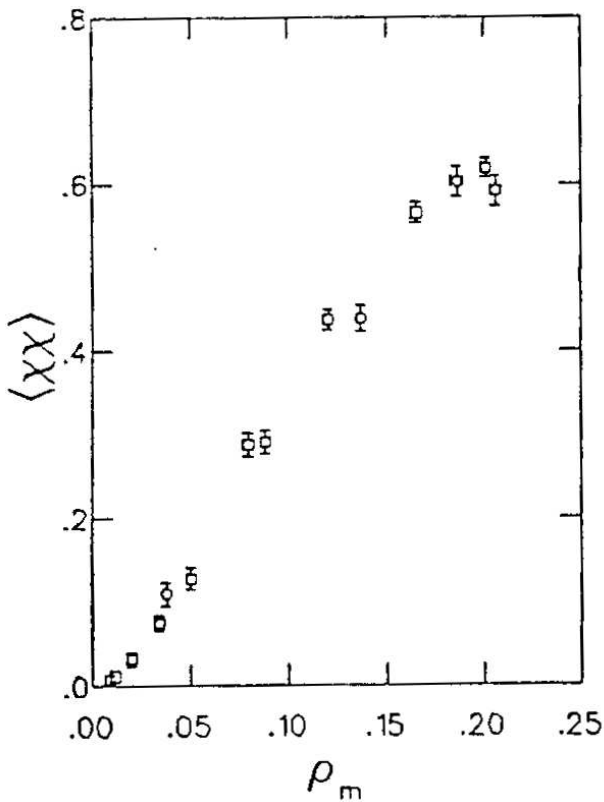}
\caption{Correlation between $\langle\overline{\chi}\chi\rangle$ and $\rho_m$
for the compact (circles) and the noncompact (boxes) theories for $N_f=2$ and
$8^3$ lattice according to Ref.~\cite{FiWo2}.}
\label{Fie_Wol}
\end{figure}

It is shown there that, when $\langle\overline{\chi}\chi\rangle$ is plotted
versus the monopole density $\rho_m$, data points for both theories fall on
the same curve to a good approximation (see Fig.~\ref{Fie_Wol}). This led the
authors of Refs.~\cite{FiWo1,FiWo2} to the conclusion that the physics of the
chiral symmetry breaking is the same in the two theories.

Our program is to study if the conclusion reached by Fiebig and
Woloshyn can be extended to the continuum limit, by looking at the
same observables they considered: namely the chiral condensate and
the monopole density.

\section{\label{sec:4} Numerical results}

Since QED$_3$ is a super-renormalizable theory, the coupling constant does
not display any lattice space dependence. The continuum limit is approached
by merely sending $\beta=1/(e^2 a)$ to infinity. In this limit all
physical quantities can be expressed in units of the scale set by the coupling
$e$. Therefore, it is natural to work in terms of dimensionless variables
such as $\beta m$, $L/\beta$ or $\beta^2 \langle\overline\chi\chi\rangle$,
which depend on $e$ ($L$ is the lattice size).

The signature that the continuum limit is approached is that data taken
at different $\beta$ should overlap on a single curve when plotted in
dimensionless units~\cite{HaKoSt}.

In practice, numerical results will not describe the correct physics
of the system even in the continuum limit because of finite volume
effects which are particularly significant in our case, due to the
presence of a massless particle, the photon. In principle one should
get rid of these effects by taking $L/\beta \rightarrow \infty$. In
practice, this ratio is taken to be large, but finite. 
In Ref.~\cite{GuRe} the authors conclude that in order to find chiral 
symmetry breaking for $N_f=2$ at least a ratio $L/\beta \approx 5 \times 10^3$
is required. In our simulations the largest value for the $L/\beta$ ratio 
has been 20.

\begin{figure}[htb]
\includegraphics[width=8.5cm]{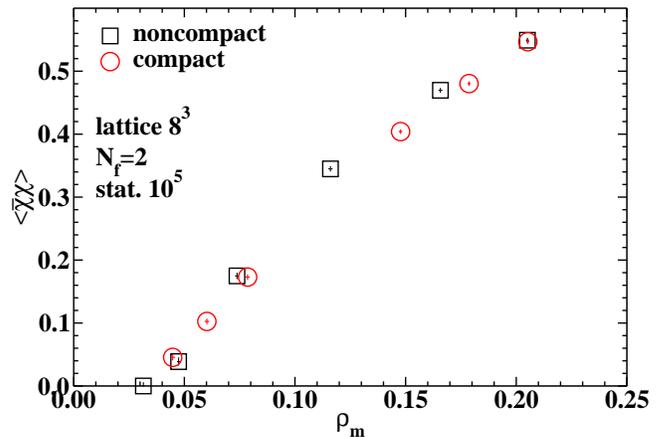}
\caption{As in Fig.~\ref{Fie_Wol}, according to our results.}
\label{ours}
\end{figure}

Our Monte Carlo simulation code was based on the hybrid updating algorithm,
with a microcanonical time step set to $dt=0.02$. We simulated one flavour of
staggered fermions corresponding to two flavours of 4-component fermions. Most
simulations were performed on a $12^3$ lattice, for bare quark mass ranging in
the interval $am=0.01\div 0.05$. We made refreshments of the gauge
(pseudofermion) fields every 7 (13) steps of the molecular dynamics. In order
to reduce autocorrelation effects, ``measurements'' were taken every
50 steps. Data were analyzed by the jackknife method combined with binning.

\begin{figure}[htb]
\includegraphics[width=8.5cm]{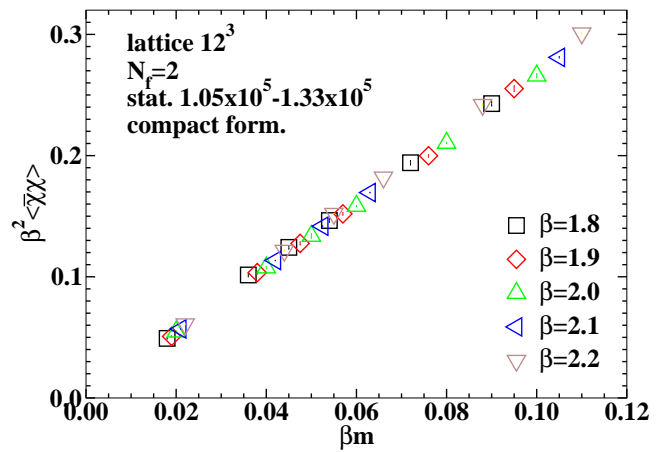}
\caption{$\beta^2 \langle\overline\chi\chi\rangle$ versus $\beta m$
in the compact formulation.}
\label{cond_comp}
\end{figure}

As a first step, we have reproduced the results by Fiebig and Woloshyn which
are shown in Fig.~\ref{Fie_Wol}. We find that also in our case data points from
the two formulations nicely overlap (see Fig.~\ref{ours}).
It should be noticed that data of Fig.~\ref{Fie_Wol} were obtained
using a linear fit with two masses ($am$=0.025, 0.05) whilst those of
Fig.~\ref{ours} have been obtained by a quadratic fit with four masses
($am$=0.02, 0.03, 0.04, 0.05), nevertheless the conclusion is the same in both
cases. 
We have verified that if we perform a linear fit on the subset of our data 
with masses $am$=0.02 and 0.05 and on the subset with masses
$am$=0.03 and 0.05, our results nicely compare with those plotted
in Fig.~\ref{Fie_Wol}.

Then, in Fig.~\ref{cond_comp} we plot data for $\beta^2
\langle\overline\chi\chi\rangle$ obtained in the compact formulation
versus $\beta m$. We restrict our attention to the subset of $\beta$
values for which data points fall approximately on the same curve,
which in the present case means $\beta=1.9, 2.0, 2.1$, corresponding
to $L/\beta=6.31,6.00, 5.71$. A linear fit of these data points
gives $\chi^2$/d.o.f. $\simeq 8.4$ and the extrapolated value for
$\beta m \rightarrow 0$ turns out to be
$\beta^2\langle\overline\chi\chi\rangle = (1.54 \pm 0.25)\times
10^{-3}$. Restricting the sample to the data at $\beta=2.1$, the
$\chi^2/$d.o.f. lowers to $\simeq 1.3$ and the extrapolated value
becomes $\beta^2\langle\overline\chi\chi\rangle = (0.94 \pm
0.28)\times 10^{-3}$, thus showing that there is a strong
instability in the determination of the chiral limit. If instead a
quadratic fit is used for the points obtained with $\beta=1.9, 2.0,
2.1$, we get $\beta^2\langle\overline\chi\chi\rangle = (0.91 \pm
0.45)\times 10^{-3}$ with
 $\chi^2$/d.o.f. $\simeq 8.7$. Owing to the large uncertainty, this
determination turns out to be compatible with both the previous ones.

\begin{figure}[htb]
\includegraphics[width=8.5cm]{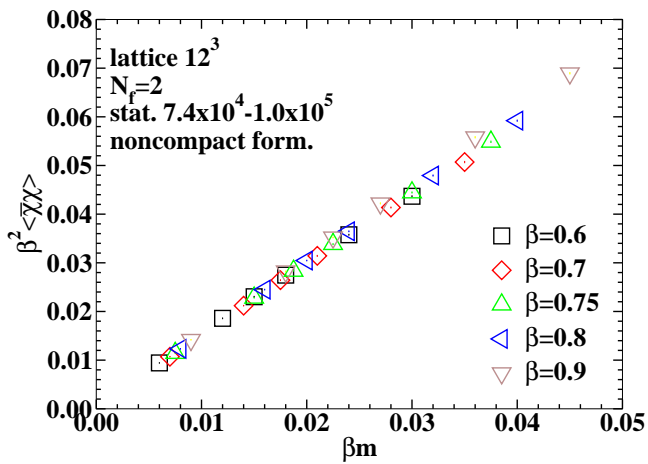}
\caption{$\beta^2 \langle\overline\chi\chi\rangle$ versus $\beta m$
in the noncompact formulation.}
\label{cond_noncomp}
\end{figure}

In Fig.~\ref{cond_noncomp} we plot data for
$\beta^2 \langle\overline\chi\chi\rangle$
obtained in the noncompact formulation versus $\beta m$. Following the
same strategy outlined before, we restrict our analysis to the data obtained
with  $\beta=0.7,0.75,0.8$, which correspond to $L/\beta= 17.14, 16, 15$.

If we consider a linear fit of these data and extrapolate to $\beta
m \rightarrow 0$, we get $\beta^2\langle\overline\chi\chi\rangle =
(0.45 \pm 0.03)\times 10^{-3}$ with $\chi^2$/d.o.f. $\simeq 17$.
Performing the fit only on the data obtained with $\beta=0.8$, for
which a linear fit gives the best $\chi^2/$d.o.f. value $\simeq 16$,
we obtain the extrapolated value
$\beta^2\langle\overline\chi\chi\rangle = (0.66 \pm 0.07)\times
10^{-3}$. Therefore, also in the noncompact formulation the chiral
extrapolation resulting from a linear fit is largely unstable. A
quadratic fit in this case gives instead a negative value for
$\beta^2\langle\overline\chi\chi\rangle$.

The comparison of the extrapolated value for
$\beta^2\langle\overline\chi\chi\rangle$ in the two formulations is
difficult owing to the instabilities of the fits and to the low
reliability of the linear fits, as suggested by the large values of
the $\chi^2/$d.o.f. Taking an optimistic point of view, one could
say that the extrapolated $\beta^2\langle\overline\chi\chi\rangle$
for $\beta=2.1$ in the compact formulation is compatible with the
extrapolated value obtained in the noncompact formulation for
$\beta=0.8$.

It is worth mentioning that our results in the noncompact formulation
are consistent with known results: indeed, if we carry out a linear fit
of the data for $\beta=0.6,0.7,0.8$ and $am$=0.02, 0.03, 0.04, 0.05 and
extrapolate, we get $\beta^2\langle\overline\chi\chi\rangle =
(1.30 \pm 0.07)\times 10^{-3}$
with an admittedly large $\chi^2/$d.o.f. $\simeq 20$, but very much in
agreement with the value $\beta^2\langle\overline\chi\chi\rangle
= (1.40 \pm 0.16)\times 10^{-3}$ obtained in Ref.~\cite{AlFaHaKoMo}.

We stress again that our results are plagued by strong finite volume
effects, therefore our conclusions on the extrapolated values of
$\beta^2\langle\overline\chi\chi\rangle$ are significant only in the
compact versus noncompact comparison we are interested in. We do not
even try to draw any conclusion from our data on the critical number
of the flavours. As a matter of fact a recent paper~\cite{HaKoSt}
shows that, if effects are carefully monitored and large lattices,
up to $50^3$, are used, it is possible to establish that
$\beta^2\langle\overline\chi\chi\rangle \leq 5 \times 10^{-5}$.
For the comparison between compact and noncompact QED$_3$ it is pertinent 
to carry out the numerical analysis with an (approximately) constant value of 
the ratio $L/\beta$.
This condition is indeed verified even if we performed simulations on lattices
with fixed ($L=12$) size, since the range of allowed values for $\beta$ 
corresponding to the continuum limit is narrow ($\beta=1.8 \div 2.2$ in
the compact case, $\beta=0.6 \div 0.9$ in the noncompact case).
Finite volume effects play a ``second order'' role in our work, since
they probably only affect the extension of the continuum limit window 
of $\beta$ values.
%

\begin{figure}[htb]
\includegraphics[width=8.5cm]{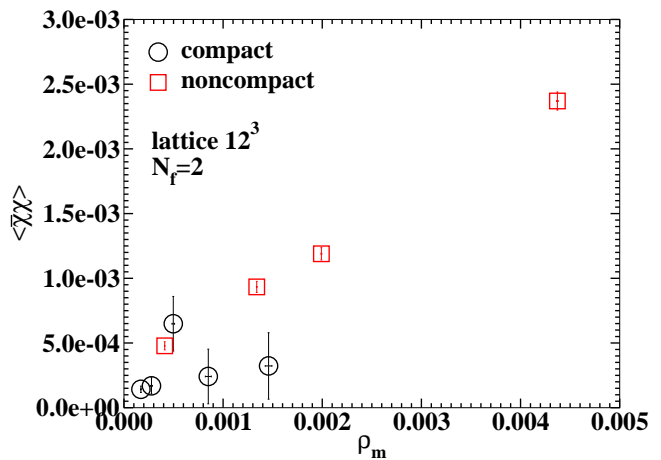}
\caption{$\langle\overline\chi\chi\rangle$ versus $\rho_m$ in both
the compact and the noncompact formulations on a $12^3$ lattice.}
\label{cond_vs_rho}
\end{figure}

In Fig.~\ref{cond_vs_rho} we plot $\langle\overline\chi\chi\rangle$ versus
the monopole density $\rho_m$. Differently from Figs.~\ref{Fie_Wol}-\ref{ours},
it is not evident with the present results that the two formulations
are equivalent also in the continuum limit, although such an equivalence
cannot yet be excluded.

\begin{figure}[htb]
\includegraphics[width=8.5cm]{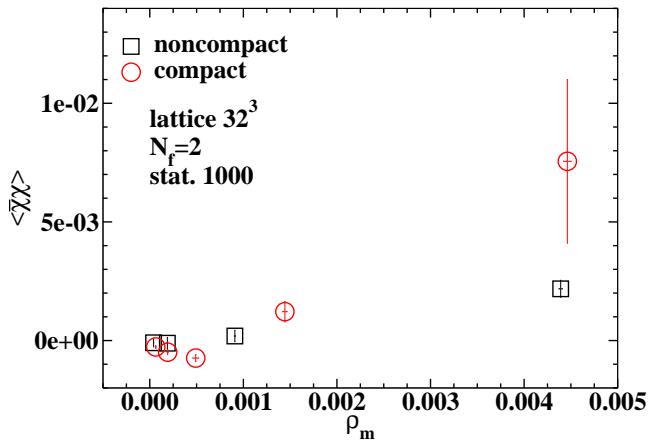}
\caption{$\langle\overline\chi\chi\rangle$ versus $\rho_m$ in both
the compact and the noncompact formulations on a $32^3$ lattice.}
\label{cond_vs_rho_2}
\end{figure}

In Fig.~\ref{cond_vs_rho_2} we plot again $\langle\overline\chi\chi\rangle$ 
versus the monopole density $\rho_m$, but now on a $32^3$ lattice.
In this case the chiral condensate is extrapolated to zero mass
by a quadratic fit. In spite of the negative value taken by 
$\langle\overline\chi\chi\rangle$ for large $\beta$,
in this case data for both formulations seem to fall on the same curve.

\begin{figure}[htb]
\includegraphics[width=8.5cm]{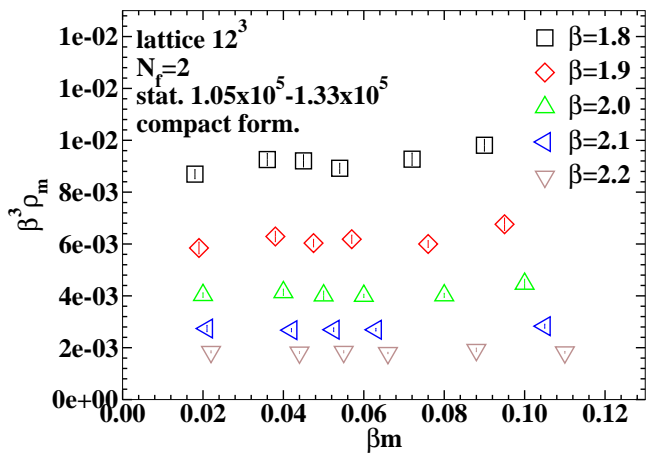}
\caption{$\beta^3\rho_m$ versus $\beta m$ in the compact formulation.}
\label{rho_comp}
\end{figure}

\begin{figure}[htb]
\includegraphics[width=8.5cm]{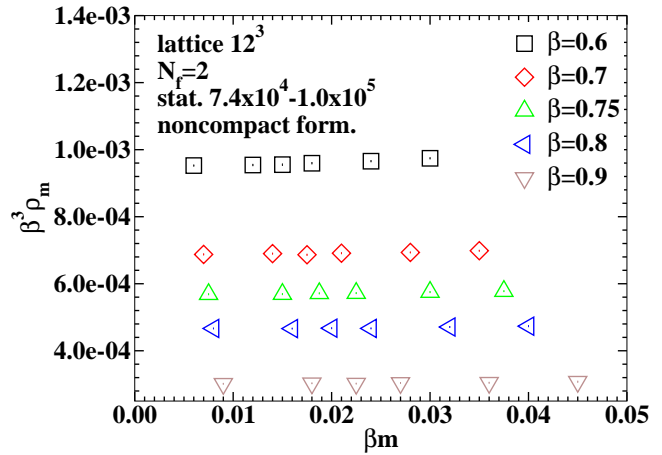}
\caption{$\beta^3\rho_m$ versus $\beta m$ in the noncompact formulation.}
\label{rho_noncomp}
\end{figure}

In Fig.~\ref{rho_comp} and Fig.~\ref{rho_noncomp} we plot $\beta^3\rho_m$
versus $\beta m$ for the two formulations; the former quantity is dimensionless,
therefore, in analogy with the previous cases, we expect that data at
different $\beta$ values should fall on a single curve in the continuum limit.
Our results show that this is not the case, this suggesting that the continuum
limit has not been reached for the monopole density.

Simulations on the $32^3$ lattice give practically the same results for
$\beta^3\rho_m$, indicating that this observable, unlike 
$\beta^2\langle\overline\chi\chi\rangle$, is volume independent. 

It is important to observe, however, that the monopole density is
independent of the fermion mass. Since the mechanism of confinement
in the theory with infinitely massive fermions, i.e. in the pure
gauge theory, is based on monopoles and since the monopole density
is not affected by the fermion mass, we may conjecture that this
same mechanism holds also in the chiral limit. This supports the
arguments by Herbut about the confinement in the presence of
massless fermion~\cite{HeSe,AzLu}. \\

\section{\label{sec:5} Conclusions}

In this paper we have compared the compact and the noncompact
formulations of QED$_3$ by looking at the behavior of the chiral
condensate and the monopole density.

Numerical results for $\beta^2 \langle\overline\chi\chi\rangle$ are compatible
with those obtained by other groups, although it is still
questionable if the continuum limit has been reached and if the
chiral limit is stable. The biggest difficulty for this observable is that the
chiral extrapolation is rough when a linear fit is performed,
but gives a negative value when instead a quadratic fit is considered.
Massive calculations on larger lattices are needed to further reduce the 
finite volume effects and to stabilize the chiral limit.

As far as monopoles are concerned, they appear in smaller and smaller numbers 
for large $\beta$, this making the determination of the continuum limit
for $\beta^3\rho_m$ rather problematic. Our results show, however, a very
weak volume dependence.

We have analyzed also the relationship between the monopole density
and the fermion mass, both in compact and noncompact QED$_3$. The
weak dependence observed leads us to conclude that the Polyakov
mechanism for confinement holds not only in the pure gauge theory,
but also in presence of massless fermions.

Finally, we note that, although the chiral condensate and monopole density
approach the continuum limit in two different ranges of $\beta$, the analysis 
{\em \`a la} ``Fiebig and Woloshin'' does not allow to exclude the equivalence of 
the compact and noncompact lattice formulations of QED$_3$.

\end{document}